\documentclass[aps,prb,reprint]{revtex4-2}

\usepackage{amsmath,amssymb,amsfonts,bm}
\usepackage{graphicx}
\usepackage{dcolumn}
\usepackage{booktabs}
\usepackage{physics}
\usepackage{hyperref}
\usepackage{xcolor}
\usepackage{amsmath}
\usepackage{amsfonts}
\usepackage{amssymb}
\usepackage{enumitem}
\usepackage{siunitx}
\usepackage{braket}
\usepackage{float}

\usepackage{subcaption}

\begin{document}

\title{Bright–Dark State Interplay Governs Vibronic Decoherence in Molecular Aggregates}

\author{Sreeja Loho Choudhury}
\affiliation{Department of Chemistry and Organic and Carbon Electronics Laboratories, North Carolina State University, 27695, USA}

\author{Abhradeep Sarkar}
\affiliation{Department of Chemistry and Organic and Carbon Electronics Laboratories, North Carolina State University, 27695, USA}

\author{Raja Ghosh}\email{rghosh8@ncsu.edu}
\affiliation{Department of Chemistry, Department of Physics, and Organic and Carbon Electronics Laboratories, North Carolina State University, 27695, USA}

\date{\today}

\begin{abstract}

In this study, we establish a spectroscopy-driven framework that connects optically bright vibronic eigenstates to their subsequent decoherence dynamics in excitonic molecular aggregates. Combining the Multiparticle Holstein Formalism with explicit system--bath interactions, we demonstrate that relaxation-induced purity loss is governed by three interconnected factors: (i) the bath spectral window, which determines the accessibility of intra- and interband relaxation pathways; (ii) the local network of dark vibronic states surrounding the optically prepared bright state; and (iii) the energetic placement of the bright state, which controls downhill relaxation. Remarkably, states with different excitation energies exhibit nearly identical purity dynamics when their local relaxation networks are similar. The resulting H--J asymmetry further reveals the importance of energetic directionality. These findings establish a reliable connection between optical spectroscopy, vibronic band structure, and state-selective loss in purity.

\end{abstract}

\maketitle

\section{Introduction}

Optical excitation provides access to bright excitonic states
whose subsequent dynamics are governed by electron--phonon
interactions and environmental fluctuations
\cite{Fuller2015ARPC,Jonas2003,Cho2008ChemRev,Schultz2024ChemRev,Scholes2017Nature,Cao2020SciAdv,BreuerPetruccione2002,Weiss08,MayKuhn2011}.
Understanding how these interactions influence quantum
coherence, energy relaxation, and dissipation remains a
central challenge.
Foundational advances have been driven by increasingly
accurate methods for treating open quantum systems,
including path-integral approaches (QuAPI)
\cite{Makri21,Makri25,Makri95I,Makri95II,Makri95III},
hierarchical equations of motion (HEOM)
\cite{TanimuraKubo1989,Tanimura2020,TanimuraIshizaki05,Tanimura06,Tanimura2014},
multiconfigurational time-dependent Hartree (MCTDH)
\cite{Irene22I,Sreejaetal26,Irene22II,Irene18I,Irene18II,Irene21},
Redfield theory \cite{David18,David17I,David17II},
stochastic Schr\"odinger equations
\cite{Zhou24,Zhao13,Zoller2004,Strunz2004},
and mixed quantum--classical methods
\cite{Grossmann20,Grossmann06,Grossmann08,Grossmann97}.
These approaches have substantially advanced our
understanding of energy and charge transfer,
nonadiabatic dynamics, and environmentally induced
decoherence in molecular systems.

Complementary studies have revealed the microscopic
mechanisms underlying molecular decoherence.
Prezhdo and Rossky connected electronic decoherence
timescales to solvent relaxation dynamics \cite{PreRos98},
while Gu and Franco demonstrated the contributions
of pure dephasing, electronic transitions, and their
interference to coherence loss \cite{GuFran18}.
Using exact quantum dynamics, Hu \textit{et al.}
established the importance of electron--nuclear
entanglement and optical state preparation \cite{Franco18}.
More recently, Gustin \textit{et al.} identified
mode-specific contributions to electronic
decoherence \cite{Francoetal23}, while Scheidegger
\textit{et al.} demonstrated the influence of
molecular size and flexibility on coherence
lifetimes \cite{PNASVani25}.
In excitonically coupled photosynthetic complexes,
Gustin \textit{et al.} identified frequency-selective
energy dissipation pathways \cite{JPCLFranco25}.
Related studies have demonstrated vibrational control
of electron-transfer pathways
(Yang \textit{et al.} \cite{Bittner17}),
mode-selective spin--phonon relaxation
(Dmitriev \textit{et al.} \cite{Bittner25}),
and the influence of bath spectral properties
on decoherence dynamics
(Gustin \textit{et al.} \cite{JPCLFranco25}).
Collectively, these studies establish that
molecular structure, nuclear dynamics, and
the availability of relaxation pathways
critically influence quantum coherence.

Building upon these advances, we address a
distinct question relevant to excitonic molecular
aggregates and $\pi$-conjugated systems:
\emph{How does the vibronic eigenstructure of
molecular aggregates govern relaxation-induced
purity loss from optically bright excitations?}
Our approach builds on the Multiparticle Holstein
Formalism (MHF), which successfully reproduces
the characteristic vibronic progressions observed
in the absorption and emission spectra of
organic molecular aggregates
\cite{ghoshspano2020,HestandSpano18,Spano10,Holstein1959I,Holstein1959II}.
The MHF explicitly captures the interplay
between intermolecular excitonic coupling
and the dominant intramolecular
aromatic-to-quinoidal stretching vibration.
Diagonalization of the system Hamiltonian
provides the energies, wavefunctions, and
oscillator strengths of vibronically dressed
eigenstates, naturally identifying the
optically bright and dark states underlying
the observed spectral features.
Rather than prescribing arbitrary electronic
initial states, we therefore initialize
the open-system dynamics in spectroscopically
identifiable bright vibronic eigenstates.
This establishes a direct connection between
optical spectroscopy and the subsequent
state-selective purity dynamics.

Molecular aggregates provide a particularly
suitable platform because the sign and
magnitude of intermolecular excitonic coupling
control both the energetic placement of
optically bright states and the surrounding
dark-state landscape. By systematically
varying excitonic coupling, including or
excluding intramolecular vibrations, and
tuning the bath spectral window, we identify
three key factors governing relaxation-induced
purity loss: (i) the bath spectral window,
which controls the accessibility of intra-
and interband relaxation pathways;
(ii) the local distribution of dark vibronic
states and their system--bath coupling to
the prepared bright state; and
(iii) the energetic placement of the bright
state, which determines the availability
of downhill relaxation channels.
In our framework, the dominant intramolecular
aromatic-to-quinoidal stretching vibration
is treated explicitly within the system
Hamiltonian, while the remaining intermolecular,
solvent, and lattice motions are represented
by an external bath responsible for relaxation
and dephasing
\cite{BreuerPetruccione2002,Weiss08,MayKuhn2011}.
Together, these results establish a
spectroscopy-driven connection between
the vibronic eigenstructure of molecular
aggregates, bath-accessible relaxation
pathways, and state-selective purity dynamics.


\section{Theoretical Model}

The total Hamiltonian is written as
\begin{equation}
H = H_S + H_B + H_{SB},
\end{equation}
where $H_S$, $H_B$, and $H_{SB}$ denote the system,
bath, and system--bath interaction Hamiltonians,
respectively. The structure of the total
system--bath Hilbert space is illustrated in
Section I of the Supplementary Information (SI).
The system Hamiltonian is described by the
Holstein Hamiltonian
\begin{align}
H_S &=
\sum_{j=1}^{J} e_j |j\rangle \langle j|
+
\sum_{j=1}^{J-1} t
\left(
|j+1\rangle\langle j|
+
\mathrm{h.c.}
\right)
\nonumber\\
&+
\hbar\omega_{\rm vib}
\sum_{j=1}^{J} b_j^\dagger b_j
+
\hbar\omega_{\rm vib}
\sum_{j=1}^{J}
\left[
\lambda
\left(
b_j^\dagger+b_j
\right)
+
\lambda^2
\right]
|j\rangle\langle j|,
\end{align}
where $|j\rangle$ denotes a Frenkel exciton
localized on site $j$, $e_j$ is the corresponding
site energy, $t$ is the intermolecular electronic
coupling, $b_j^\dagger$ creates an intramolecular
vibrational excitation of frequency
$\omega_{\rm vib}$, and $\lambda^2$ is the
Huang--Rhys factor. Throughout this work, we set $\hbar\omega_=0.174$, corresponding to the aromatic-to-quinoidal stretching mode in $\pi$-conjugated systems, and $\lambda^2=1$, a value typical of Frenkel excitons \cite{paquin2013two,ghoshspano2020,Spano10}.
Two coupling strengths are considered,
$t=\pm0.05~{\rm eV}$ and $t=\pm0.4~{\rm eV}$,
corresponding to weak- and strong-excitonic-coupling
regimes, respectively. The eigenstates of $H_S$ are obtained using the multiparticle basis set
\cite{ghoshspano2020,HestandSpano18},
\begin{align}
|\psi_i\rangle
&=
\sum_{j,\tilde v}
c_{j,\tilde v}^{(i)}
|j,\tilde v\rangle
\nonumber\\
&+
\sum_{j,\tilde v}
\sum_{j',v'}
c_{j,\tilde v,j',v'}^{(i)}
|j,\tilde v;j',v'\rangle,
\end{align}
where $|j,\tilde v\rangle$ and
$|j,\tilde v;j',v'\rangle$ represent
one-particle and two-particle basis states,
respectively. In the one-particle basis,
an electronically excited chromophore at
site $j$ carries $\tilde v$ vibrational quanta
in the excited-state nuclear potential,
while all other sites remain electronically
and vibrationally unexcited. In the
two-particle basis, an electronically excited
chromophore at site $j$ carries $\tilde v$
vibrational quanta, while a second,
electronically unexcited site $j'\neq j$
carries $v'\geq1$ vibrational quanta in
the ground-state nuclear potential.
An illustration of the construction of
the multiparticle basis states for a given
vibrational truncation is provided in the SI.

Diagonalization of $H_S$ using this multiparticle basis yields vibronically
dressed excitonic eigenstates and their
corresponding energies, $E_i$. Within the electric-dipole
approximation, the absorption spectrum is
calculated as
\begin{equation}
A(\omega)\propto
\sum_i
\left|
\langle\psi_i|\hat{\boldsymbol{\mu}}|G\rangle
\right|^2
L(E_i-E_G-\hbar\omega),
\end{equation}
where $|G\rangle$ is the electronic and
vibrational ground state, $E_G$ is its
energy, $\hat{\boldsymbol{\mu}}$ is the
transition dipole operator, and $L$ is
a normalized lineshape function. The
transition strength of each vibronic
eigenstate is determined by
\begin{equation}
I_i=
\left|
\langle\psi_i|\hat{\boldsymbol{\mu}}|G\rangle
\right|^2.
\end{equation}
Consequently, eigenstates with nonzero
transition strengths contribute to the
absorption spectrum and are identified
as optically bright states, whereas
states with vanishing transition strengths
are optically dark. The interplay between
intermolecular electronic coupling and
intramolecular vibrations determines the
energies, vibronic character, and oscillator
strength distribution of these eigenstates.

To describe the subsequent open-system
dynamics, the environment is modeled as
a collection of harmonic oscillators,
\begin{equation}
H_B
=
\sum_{n=1}^{N}
\hbar\omega_n
a_n^\dagger a_n,
\end{equation}
coupled locally to the excitonic subsystem
through
\begin{equation}
H_{SB}
=
\sum_{n,j}
g_{n,j}
\left(
a_n^\dagger+a_n
\right)
|j\rangle\langle j|.
\end{equation}
To preserve the symmetry of the dimer
while generating decoherence, neighboring
sites are assigned opposite coupling phases,
\begin{equation}
g_{n,j}=s_j g_n,
\qquad
s_j=(-1)^{j+1}.
\end{equation}
In the purely electronic limit, this
interaction reduces to a $\sigma_z$-type
system--bath coupling. The bath modes are chosen to discretize
an Ohmic spectral density,
\begin{equation}
J(\omega)
=
2\alpha
\omega_c^{\,1-s}
\omega^{s}
e^{-\omega/\omega_c},
\end{equation}
where $\alpha$ is the dimensionless
system--bath coupling strength, $\omega_c$
is the bath cutoff frequency, and $s=1$
corresponds to an Ohmic environment
\cite{Leggett87,Weiss08}. Since explicit
propagation requires a finite number of
bath modes, the continuum is discretized
using logarithmic frequency sampling
\cite{Bulla03,Chin10,DPM20},
\begin{equation}
\omega_n
=
-\omega_c
\ln
\left(
1-\frac{n}{N_c}
\right),
\qquad
n=1,2,\ldots,N,
\end{equation}
where $N_c=N+\Delta N$. Here, $\Delta N$
is a numerical regularization parameter
introduced to prevent the logarithmic
discretization from sampling the singular
endpoint $n=N_c$. Throughout this work,
we choose $\Delta N=10$, which provides
a finite high-frequency cutoff while
ensuring numerical convergence of the
reported results. The corresponding
system--bath couplings are
\begin{equation}
g_n
=
\sqrt{
\alpha
\frac{\omega_c^{\,2-s}}{N_c}
\omega_n^{\,s}
}.
\end{equation}

The full system--bath wavefunction is
propagated according to
\begin{equation}
|\Psi(t)\rangle
=
e^{-iHt/\hbar}
|\Psi(0)\rangle,
\end{equation}
with an initially factorized state,
\begin{equation}
|\Psi(0)\rangle
=
|\psi_{\rm sys}(0)\rangle
\otimes
|0\rangle_{\rm bath},
\end{equation}
where $|\psi_{\rm sys}(0)\rangle$ is the
selected optically bright eigenstate of
$H_S$ and $|0\rangle_{\rm bath}$ denotes
the bath vacuum, corresponding to an
initially zero-temperature environment
\cite{Feynman98}. Decoherence is quantified using the
reduced density matrix \cite{Schloss},
\begin{equation}
\rho_{\rm red}(t)
=
{\rm Tr}_B
\left[
|\Psi(t)\rangle
\langle\Psi(t)|
\right],
\end{equation}
and the subsystem purity,
\begin{equation}
\mathcal P(t)
=
{\rm Tr}
\left[
\rho_{\rm red}^2(t)
\right].
\end{equation}
The purity equals unity for a pure
subsystem state and decreases as
system--bath entanglement develops.
Throughout this work, purity serves
as the primary measure of decoherence,
enabling a direct comparison of how
the optically prepared bright state,
its surrounding vibronic eigenstate
landscape, and the environmental
spectral window jointly govern
coherence loss.

\begin{figure*}
\includegraphics[width=\linewidth]
{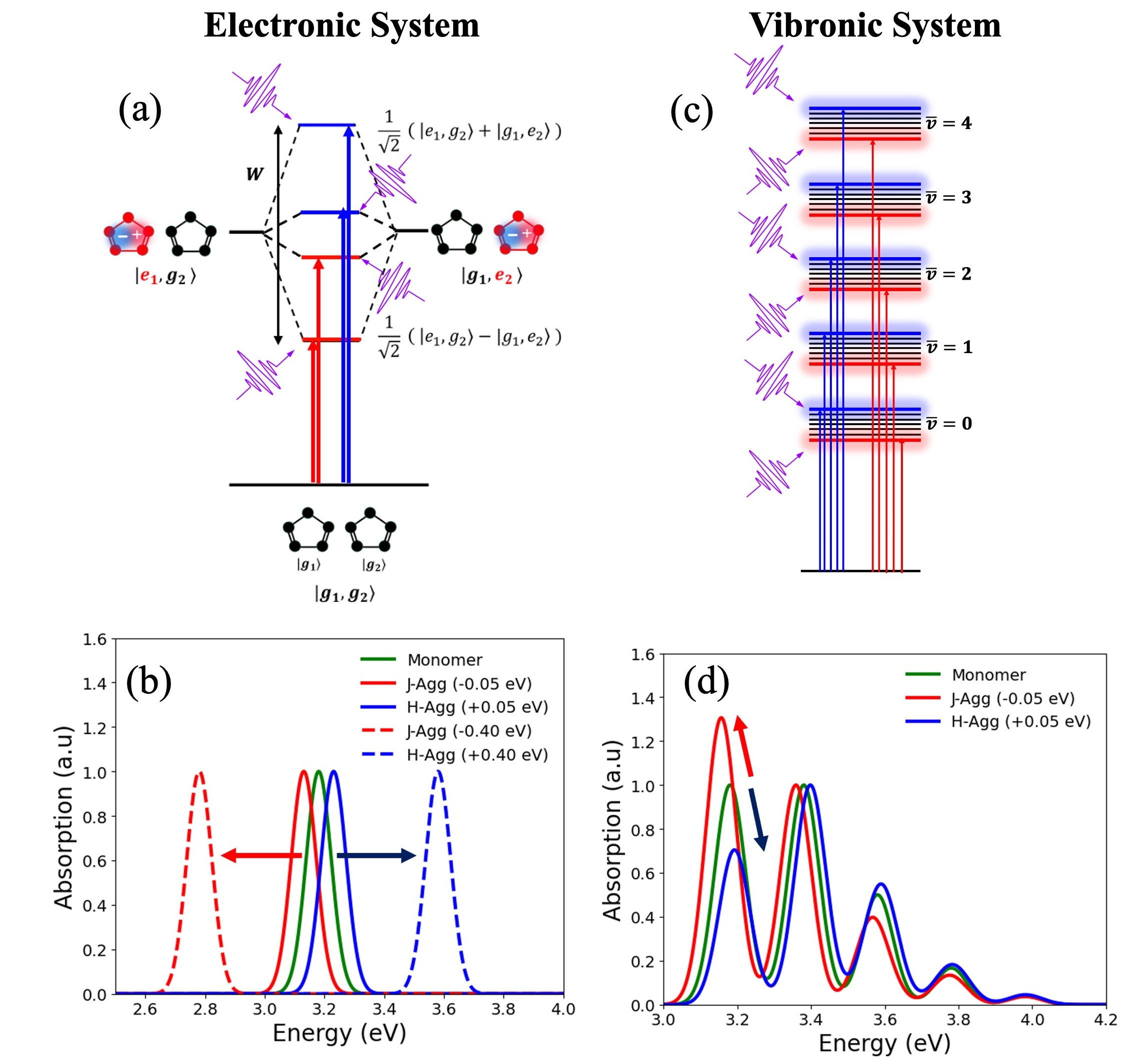}
\caption{\label{fig:0}
(a) Excitonic splitting and optical selection
rules in the purely electronic limit.
(b) Redshifted (J-aggregate, $t<0$) and
blueshifted (H-aggregate, $t>0$) absorption
relative to the monomer for weak
($|t|=0.05$ eV) and strong
($|t|=0.4$ eV) electronic coupling.
(c) Vibronically dressed eigenstates
organized into manifolds ($\bar v=0$--$4$),
illustrating optically bright and dark states.
(d) Vibronic absorption spectra showing
the redistribution of oscillator strength
among the $0$--$0$, $0$--$1$, and
higher-order transitions for H- and
J-aggregates.
}
\end{figure*}

\section{Results and Discussion}
\label{sec:results}

We first validate our numerical approach against the
analytically solvable pure-dephasing limit ($t=0$),
in which a two-level system initially prepared in a
coherent superposition interacts with a harmonic
bath \cite{Schloss}. As detailed in Section III of
the Supplementary Information (SI), the numerical
purity dynamics agree excellently with the analytical
solution for both the purely electronic and vibronic
dimers. In Section IV of the SI, we further extend
the analytical decoherence factor to larger molecular
aggregates and multiparticle Holstein basis sets
incorporating intramolecular vibrations. These
benchmarks establish the reliability of our numerical
approach for investigating decoherence in electronically
coupled molecular systems.

\subsection{Optically Bright and Dark Vibronic Eigenstates}

To connect the system Hamiltonian with optically
accessible eigenstates, we first consider the purely
electronic dimer, neglecting intramolecular vibrations.
As illustrated in Fig.~\ref{fig:0}(a), intermolecular
electronic coupling splits the degenerate monomer
excitation into two excitonic eigenstates separated
by bandwidth ($W=2|t|$). The sign of $t$ determines the position of the totally symmetric state and selection rules.
For J-aggregates ($t<0$), the lower-energy
excitonic state is optically bright, producing
a redshifted absorption relative to the monomer.
Conversely, for H-aggregates ($t>0$), the
upper-energy state is bright, resulting in
a blueshifted absorption
[Fig.~\ref{fig:0}(b)]. 

Including intramolecular vibrational coupling
generates vibronically
dressed eigenstates organized into different manifolds
[Fig.~\ref{fig:0}(c)]. Within each vibronic manifold,
intermolecular electronic coupling redistributes
oscillator strength among the eigenstates,
producing optically bright and dark states.
Consequently, the absorption spectrum develops
a vibronic progression comprising the
$0$--$0$, $0$--$1$, $0$--$2$, and higher-order
transitions, with distinct intensity distributions
for H- and J-aggregates
[Fig.~\ref{fig:0}(c)]. Neverethless, the top (bottom) state within each vibronic manifold is optically bright. 

Importantly, the accessibility of the bright and dark
states emerges naturally from diagonalization
of the system Hamiltonian and evaluation of
their transition dipole strengths, rather than
being assigned \emph{a priori}. This establishes
a direct connection between experimentally
observable absorption signatures and the
underlying vibronic eigenstate landscape,
providing the foundation for selecting these
optically bright initial states and examining
their subsequent decoherence dynamics.

\begin{figure*}
\includegraphics[width=\linewidth]
{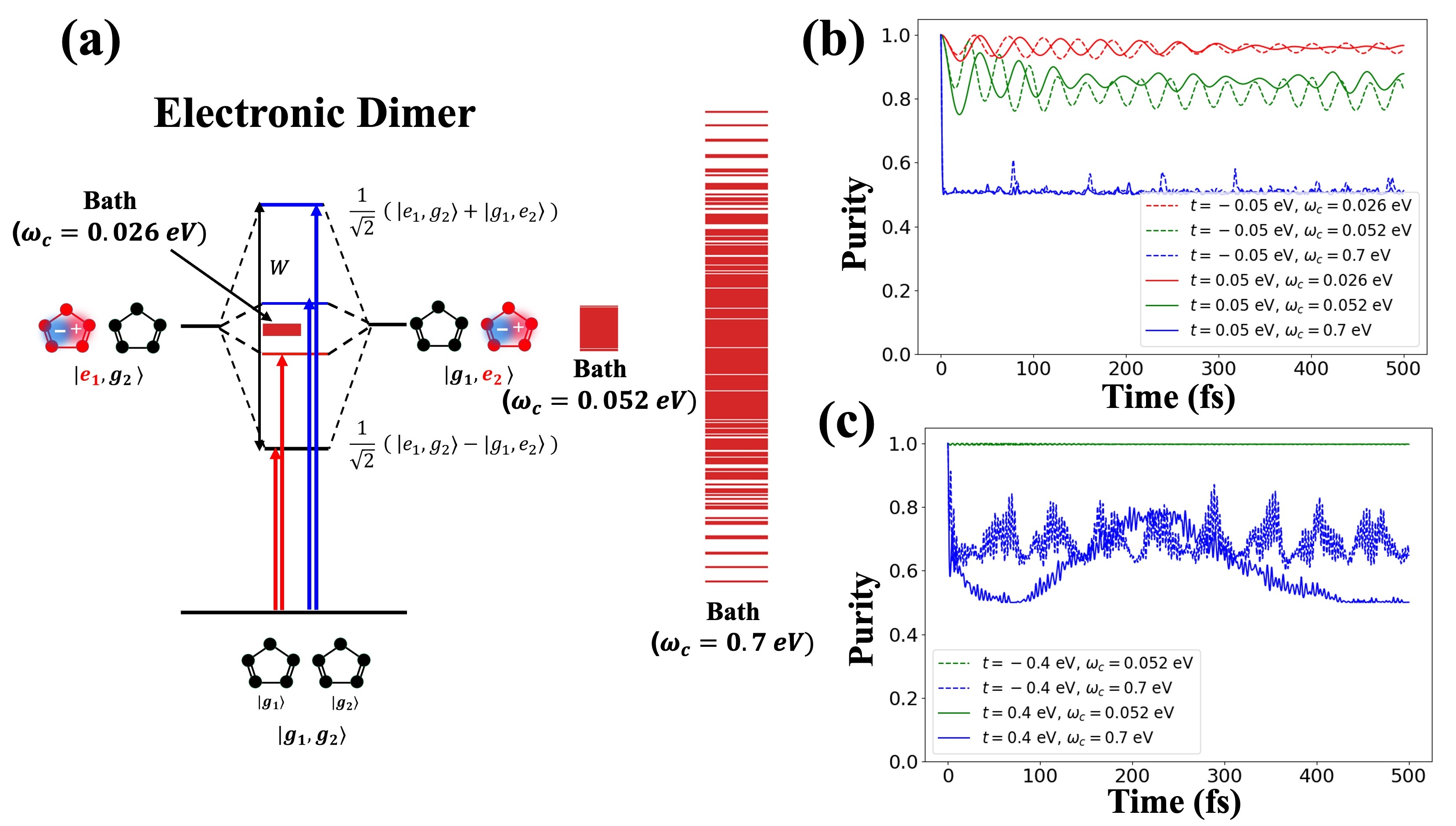}
\caption{\label{fig:1}
(a) Schematic of the purely electronic dimer
illustrating the optically bright excitonic
states and three Ohmic bath spectral windows:
off-resonant ($\omega_c=0.026$ eV),
near-resonant ($\omega_c=0.052$ eV),
and broad-band ($\omega_c=0.7$ eV).
(b) Purity dynamics in the weak excitonic-coupling
regime ($t=\pm0.05$ eV).
(c) Purity dynamics in the strong excitonic-coupling
regime ($t=\pm0.4$ eV).
}
\end{figure*}
\subsection{Bath Spectral Control of Electronic Decoherence}

To
isolate the influence of the bath spectral
window on excitonic decoherence, we start the analysis with the purely electronic dimer. 
Unless otherwise stated, the calculations
employ five bath modes with three Fock states
per mode. Convergence with respect to bath
discretization and truncation is demonstrated
in Section V of the SI. 

Bath-induced relaxation is controlled by the
spectral overlap between the environment and
the electronic energy splitting,
$\Delta E_{12}=W$. Efficient transitions
require appreciable bath spectral weight
at the corresponding transition frequency.
We distinguish three bath spectral window regimes:
(i) off-resonant ($\omega_c=0.026$ eV), where the bath has negligible
spectral weight at $W$;
(ii) near-resonant ($\omega_c=0.052$ eV), where appreciable spectral
weight overlaps the excitonic splitting;
and (iii) broad-band($\omega_c=0.7$ eV), where the environment
supports a wider range of transition energies.

Figure~\ref{fig:1}(b) compares the purity dynamics in the weak excitonic-coupling regime ($t=\pm0.05$ eV, $W=0.1$ eV). Regardless of the choice of initial state, corresponding to the optically bright excitonic state (upper or lower depending on the sign of $t$), relaxation is strongly suppressed for the off-resonant bath ($\omega_c=0.026$ eV, red curves). Consequently, the subsystem remains nearly pure, exhibiting only weak oscillatory modulations. Increasing the bath cutoff to $\omega_c=0.052$ eV enhances the spectral overlap with the excitonic splitting, resulting in a greater loss of purity accompanied by coherent oscillations (green curves). For the broadband bath ($\omega_c=0.7$ eV), the environment efficiently facilitates transitions between the electronic eigenstates, leading to rapid decoherence and a substantial reduction in purity [Fig.~\ref{fig:1}(b)].

Increasing the excitonic coupling to $t=\pm0.4$ eV enlarges the electronic splitting to $W=0.8$ eV. Consequently, both lower-frequency baths become effectively off-resonant, and the purity remains close to unity, as shown by the green curves in Fig.~\ref{fig:1}(c). Appreciable decoherence occurs only when the bath spectrum extends to energies comparable to the enlarged excitonic splitting, as observed for the broadband environment. These results demonstrate that electronic decoherence is governed not simply by the strength of intermolecular excitonic coupling, but more fundamentally by the spectral accessibility of bath-induced transitions between electronic eigenstates.

\begin{figure*}
\includegraphics[width=\linewidth]
{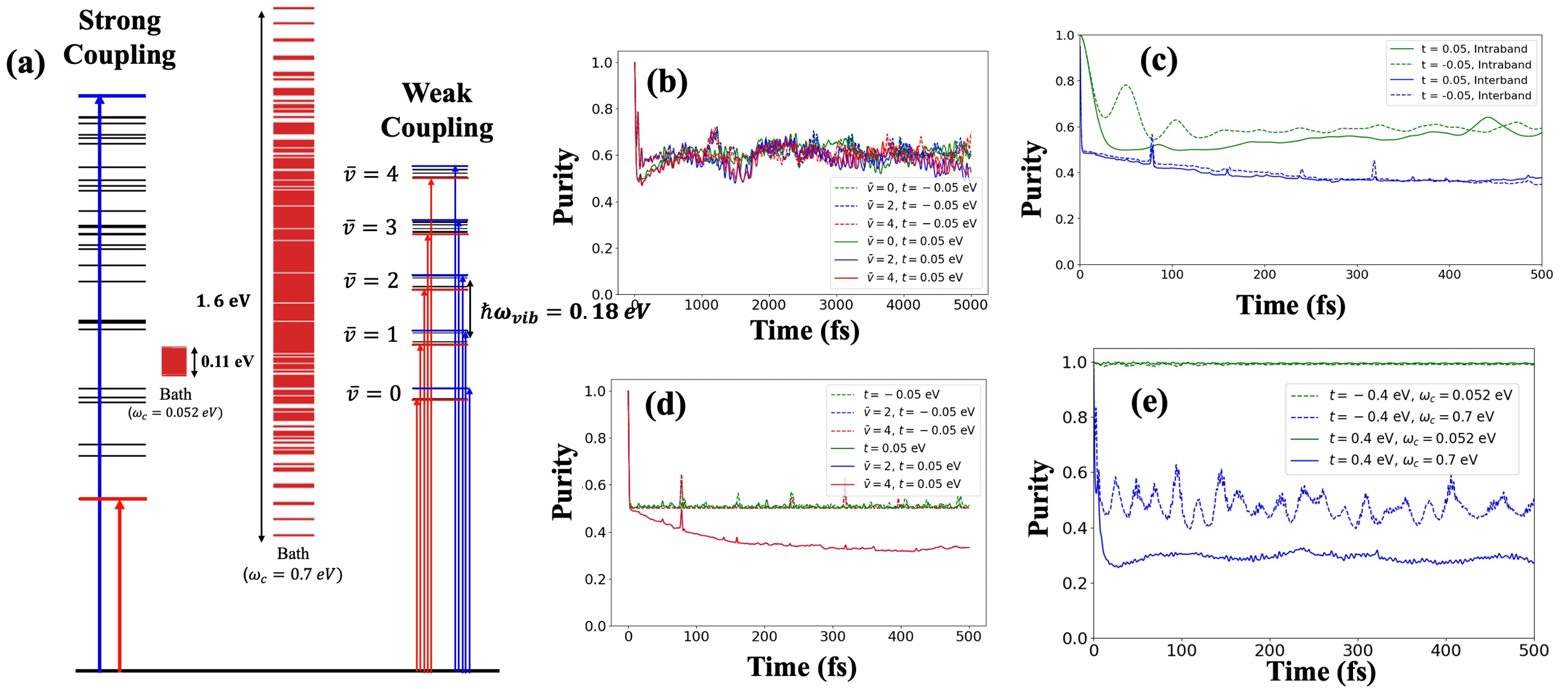}
\caption{\label{fig:2}
(a) Schematic comparison of the weak
($t=\pm0.05$ eV) and strong
($t=\pm0.4$ eV) excitonic-coupling
regimes. For $W<\hbar\omega_{\rm vib}$,
the spectrum exhibits approximately
resolved vibronic manifolds
($\bar v=0,1,\ldots,4$), whereas
for $W>\hbar\omega_{\rm vib}$,
the manifolds hybridize into mixed
vibronic states. The low-frequency
bath ($\omega_c=0.052$ eV)
predominantly probes intraband
transitions, while the broad-band
bath ($\omega_c=0.7$ eV)
additionally activates interband
relaxation pathways.
(b) Purity dynamics following selective excitation of the optically bright states in the $\bar{v}=0$, $2$, and $4$ manifolds in the weak-coupling regime for the low-frequency bath, shown up to $5000$ fs.
(c) Comparison of the purity dynamics for a vibronic dimer in the weak excitonic coupling regime, initialized in the $\bar{v}=0$ bright state, under low- and high-frequency bath conditions.
(d) Comparison of the purity dynamics for dimers with maximum vibrational truncations of $\bar{v}=0$, $\bar{v}=2$ and $\bar{v}=4$ in the weak-coupling regime, initialized in the optically bright states of the $\bar{v}=0, 2, 4$ manifolds respectively, under broad-band bath.
(e) Purity dynamics in the strong
excitonic-coupling regime under
low- and high-frequency baths.}
\end{figure*}
\subsection{State-Selective Vibronic Decoherence: Intraband vs Interband}

The incorporation of vibronic coupling changes the system from a simple two-level system to a richer vibronic
bandstructure as shown in Fig.~\ref{fig:2}a. 
In the weak excitonic-coupling regime
($W<\hbar\omega_{\rm vib}$), the spectrum
consists of vibronic
manifolds, $\bar v=0,1,2,\ldots$, separated
by the vibrational energy
$\hbar\omega_{\rm vib}=0.174$ eV
[Fig.~\ref{fig:2}(a)].
The optically bright states within each vibronic manifold (either top or bottom state) provide distinct initial conditions
for probing state-selective decoherence.

Bath-induced transitions satisfy the same spectral-overlap criterion introduced in the previous section, but may now occur either within a vibronic manifold (intraband) or between different manifolds (interband). The available relaxation pathways are again determined by the relative magnitudes of the bath spectral window and the intramolecular vibrational frequency, $\omega_{\rm vib}$,
\begin{align}
\omega_{\rm bath}^{\rm max}<\hbar\omega_{\rm vib},
&&\text{intraband decoherence},
\\
\omega_{\rm bath}^{\rm max}\ge\hbar\omega_{\rm vib},
&&\text{intra- and interband decoherence}.
\end{align}

Figure~\ref{fig:2}(b) illustrates the regime of intraband decoherence, where $\omega_{\rm bath}^{\rm max}<\hbar\omega_{\rm vib}$, such that the bath cannot bridge adjacent vibronic manifolds. Consequently, only transitions within the selected vibronic manifold contribute to decoherence. In Fig.~\ref{fig:2}(b), the upper and lower bright states correspond to negative and positive excitonic couplings, respectively. The long-time purity dynamics obtained by selectively exciting the bright states of the $\bar{v}=0$, $\bar{v}=2$, and $\bar{v}=4$ manifolds are similar. Because the bath frequencies probe only intraband energy scales, the purity decay is insensitive to the absolute excitation energy and instead depends solely on the local band structure surrounding the selected initial bright state within the corresponding vibronic manifold.

The situation changes qualitatively once the bath spectrum is broadened such that $\omega_{\rm bath}^{\rm max}>\hbar\omega_{\rm vib}$, allowing the bath to acquire sufficient spectral weight to bridge adjacent vibronic manifolds. To demonstrate the combined effects of intra- and interband decoherence, Fig.~\ref{fig:2}(c) compares the purity dynamics following excitation of the bright state (either the upper or lower bright state) in the $\bar{v}=0$ manifold of a system with a maximum vibrational truncation of $\bar{v}=4$, as shown in Fig.~\ref{fig:2}(a), for the two cases: $\omega_{\rm bath}^{\rm max}<\hbar\omega_{\rm vib}$ and $\omega_{\rm bath}^{\rm max}>\hbar\omega_{\rm vib}$. When $\omega_{\rm bath}^{\rm max}>\hbar\omega_{\rm vib}$, the bath activates additional interband relaxation pathways, therefore increasing the number of energetically accessible dark states and leading to a significantly greater loss of purity (blue curves in Fig.~\ref{fig:2}(c)) compared to $\omega_{\rm bath}^{\rm max}>\hbar\omega_{\rm vib}$ (green curves in Fig.~\ref{fig:2}(c)). The bath therefore acts as a spectral filter, determining whether decoherence remains confined to a single vibronic manifold or is distributed across multiple vibronic manifolds.

To distinguish the role of the absolute excitation energy from that of the surrounding spectral landscape, we next compare two systems with different maximum vibrational truncations: one with $\bar{v}=2$ and the other with $\bar{v}=4$. In both cases, the dynamics are initiated from the optically bright state in the highest vibronic manifold (i.e., $\bar{v}=2$ and $\bar{v}=4$, respectively) under conditions where the bath can induce interband relaxation, $\omega_{\rm bath}^{\rm max}>\hbar\omega_{\rm vib}$. Although these two initial bright states (either the upper or lower bright state) differ in energy by approximately $2\hbar\omega_{\rm vib}$, their local distributions of nearby dark states and energetically accessible relaxation pathways are nearly identical. Consequently, the corresponding purity dynamics overlap almost perfectly, as shown in Fig.~\ref{fig:2}(d). 

These results identify the local
relaxation network surrounding an
initial vibronic eigenstate $i$
as a key determinant of decoherence:
\begin{equation}
\mathcal{N}_i
=
\left\{
\Delta E_{ij},
|V_{ij}|^2,
J(\Delta E_{ij})
\right\}_{j\neq i},
\qquad
\Delta E_{ij}=|E_i-E_j|.
\label{eq26}
\end{equation}
Here,
\begin{equation}
V_{ij}
=
\langle\psi_j|\hat S|\psi_i\rangle,
\qquad
\hat S=
\sum_k s_k|k\rangle\langle k|,
\end{equation}
is the system--bath coupling matrix
element connecting vibronic eigenstates
$i$ and $j$. The quantity
$|V_{ij}|^2J(\Delta E_{ij})$
characterizes the coupling strength
and spectral accessibility of each
relaxation pathway. Thus, states
with different absolute excitation
energies can exhibit nearly identical
decoherence dynamics when their
local relaxation networks are similar.

\subsection{Energetic Directionality and H--J Asymmetry}

Despite the similarities between
the local vibronic relaxation networks,
the purity dynamics exhibit a
systematic asymmetry between H- and
J-dimers. This behavior can be
understood by considering the
directionality of bath-assisted
energy exchange at zero temperature.

For an optically bright state
$|B\rangle$ lying above a dark state
$|D\rangle$, with
$\Delta E=E_B-E_D>0$,
downhill relaxation transfers energy
from the system to the bath:
\begin{equation}
|B\rangle|n\rangle_B
\rightarrow
|D\rangle|n+1\rangle_B.
\end{equation}
The corresponding transition rate
scales as
\begin{equation}
\Gamma_{\rm down}
\propto
|V_{DB}|^2
J(\Delta E)
[n(\Delta E)+1],
\end{equation}
where $n(\Delta E)$ is the
Bose--Einstein occupation factor. Conversely, excitation from a
lower-energy bright state to a
higher-energy dark state requires
energy absorption from the bath:
\begin{equation}
|B\rangle|n\rangle_B
\rightarrow
|D\rangle|n-1\rangle_B,
\end{equation}
with
\begin{equation}
\Gamma_{\rm up}
\propto
|V_{DB}|^2
J(\Delta E)n(\Delta E).
\end{equation}
For the initially zero-temperature
bath used here, $n(\Delta E)=0$,
giving
\begin{equation}
\Gamma_{\rm up}=0,
\qquad
\Gamma_{\rm down}
\propto
|V_{DB}|^2J(\Delta E).
\end{equation}

Thus, an upper bright state can
undergo energetically downhill
relaxation by transferring energy
to the bath, whereas excitation
from a lower bright state to
higher-lying dark states is
suppressed at zero temperature.
This directional asymmetry provides
a physical explanation for the
enhanced purity loss of upper
bright states relative to lower
bright states. The H--J difference
therefore reflects not only the
distribution of accessible vibronic
states but also the energetic
position of the optically prepared
state within that distribution.

Finally, we examine the strong
excitonic-coupling regime
($t=\pm0.4$ eV), where
$W>\hbar\omega_{\rm vib}$.
Here, the electronic bandwidth
exceeds the vibrational spacing,
and adjacent vibronic manifolds
hybridize into a dense spectrum
of mixed vibronic states
[Fig.~\ref{fig:2}(a)].
Although the individual manifolds
are no longer well resolved,
optically bright eigenstates
remain identifiable through
their transition dipole strengths.

For the low-frequency bath,
relaxation remains strongly
restricted and the subsystem
retains high purity
[Fig.~\ref{fig:2}(e)].
In contrast, the broad-band
environment activates a larger
network of vibronic transitions,
producing substantial purity loss.
Notably, the H--J asymmetry
persists even in this strongly
hybridized regime, as shown by
the solid and dotted blue curves
in Fig.~\ref{fig:2}(e).
This demonstrates that the
energetic position of the
optically prepared bright state
remains an important determinant
of decoherence even when distinct
vibronic manifolds cease to exist.

\subsection{Going Beyond Dimers: Larger Molecular Aggregates}

To examine whether the trends identified for dimers extend to larger molecular aggregates, we consider a five-site excitonic chain (pentamer) in the weak-coupling regime ($t=\pm 0.05$ eV). As shown in Fig.~5 of the SI, the absorption spectra retain the characteristic H- and J-aggregate signatures, with enhanced (suppressed) 0--0 absorption for J-aggregates (H-aggregates) relative to the monomer. Figure~6 of the SI compares the purity dynamics following excitation of the optically bright states under low-frequency ($\omega_c=0.052$ eV) and broad-band ($\omega_c=0.7$ eV) environments. The low-frequency bath produces only modest purity loss, whereas the broad-band bath leads to substantially stronger decoherence for both H- and J-aggregates. These results demonstrate that bath spectral selectivity remains an important determinant of decoherence beyond dimers. While the present pentamer calculations are restricted to the purely electronic limit, future studies will incorporate intramolecular vibronic coupling to investigate how the increasing density of vibronically dressed states and the resulting relaxation pathways influence state-selective decoherence in larger molecular aggregates.

\section{Conclusion}

A central finding is that decoherence depends primarily on the local relaxation network rather than the absolute excitation energy. Remarkably, bright states separated by several vibrational quanta exhibit nearly identical purity dynamics when their surrounding relaxation networks are similar. Furthermore, the energetic directionality of bath-induced transitions explains the pronounced H--J asymmetry, which persists even when strong excitonic coupling hybridizes distinct vibronic manifolds. Preliminary calculations on larger aggregates confirm that bath spectral selectivity remains important beyond dimers, motivating future investigations incorporating vibronic coupling and finite-temperature effects.

Our results suggest three complementary strategies for tuning coherence decay. First, the bath spectral window acts as a tunable spectral filter that determines which relaxation pathways are energetically accessible. Restricting the bath frequencies below the vibrational quantum suppresses interband relaxation and confines decoherence to intraband processes, whereas extending the bath spectrum activates additional interband channels and accelerates coherence loss. Second, the local spectral environment surrounding the optically prepared bright state, rather than its absolute excitation energy, dictates the decoherence rate. Bright states embedded within similar local distributions of dark states exhibit nearly identical coherence dynamics even when separated by several vibrational quanta. Finally, the choice of the initially prepared quantum state provides an additional means of controlling decoherence. Optically preparing the upper bright state favors energetically downhill relaxation into the environment, whereas preparing the lower bright state suppresses relaxation because it requires bath-assisted uphill transitions. Consequently, H-aggregates exhibit systematically stronger decoherence than their J-aggregate counterparts despite possessing identical Hamiltonians and local bright--dark couplings.

The present analysis addresses a specific regime of excitonic decoherence: an initially prepared optically bright state at zero temperature, for which relaxation through energetically accessible vibronic states constitutes the dominant mechanism of coherence loss. More general initial conditions can introduce additional contributions. In particular, coherent superpositions of energy eigenstates can undergo pure dephasing even in the absence of population relaxation, while at finite temperature both upward and downward bath-induced transitions become possible. Optical preparation in the presence of the environment may also generate system–bath correlations that are absent from the initially factorized states considered here. The present results should therefore be interpreted as identifying the microscopic origin of state-selective relaxation-induced decoherence in this low-temperature regime, rather than as a general description of all decoherence mechanisms in molecular excitonic systems.

These findings demonstrate that the fate of an optically prepared excitation is determined not simply by its energy or excitonic coupling strength, but by the surrounding landscape of bath-accessible vibronic states. By linking experimentally observable optical signatures to microscopic relaxation pathways, our work provides a foundation for understanding and engineering state-selective decoherence in molecular materials.

\bibliography{references}

\end{document}



\title{Supplemental Material:\\
Bright–Dark State Interplay Governs Vibronic Decoherence in Molecular Aggregates}

\author{Sreeja Loho Choudhury}
\affiliation{Department of Chemistry and Organic and Carbon Electronics Laboratories, North Carolina State University, 27695, USA}

\author{Abhradeep Sarkar}
\affiliation{Department of Chemistry and Organic and Carbon Electronics Laboratories, North Carolina State University, 27695, USA}

\author{Raja Ghosh}\email{rghosh8@ncsu.edu}
\affiliation{Department of Chemistry, Department of Physics, and Organic and Carbon Electronics Laboratories, North Carolina State University, 27695, USA}
\maketitle


\section{Dimension of Hilbert Space}

In this section we provide the Hilbert space dimension of the total Hamiltonian consisting of the multiparticle Holstein Hamiltonian \cite{ghoshspano2020,ghosh2016polaron,HestandSpano18}, bath Hamiltonian and system-bath interaction term given by 

\begin{align}
H
=
H_{S}
+
H_{B}
+
H_{SB}.
\end{align}
where 
\begin{align}
H_S &=
\sum_{j=1}^{J} e_j |j\rangle \langle j|
+
\sum_{j=1}^{J-1} t
\left(
|j+1\rangle\langle j|
+
\mathrm{h.c.}
\right)
\nonumber\\
&+
\hbar\omega_{\rm vib}
\sum_{j=1}^{J} b_j^\dagger b_j
+
\hbar\omega_{\rm vib}
\sum_{j=1}^{J}
\left[
\lambda
\left(
b_j^\dagger+b_j
\right)
+
\lambda^2
\right]
|j\rangle\langle j|
\end{align}
The total Hilbert space can be expressed as $H_{S}\bigotimes H_{B}$. The bath Hamiltonian has a dimension of $M^{N}$, where $M$ is the number of fock basis states used to represent each harmonic oscillator bath. Hence, the total dimension of the Hilbert space is $J\times M^{N}$, where $J$ is the number of chromophores or sites when we consider no intramolecular vibrations. The moment we consider vibration on the individual sites, then the total Hilbert space becomes: number of multiparticle basis states $\times M^{N}$. 

In this work, we consider until two-particle basis states in presence of intramolecular vibrations. 

\section{Multiparticle basis states}
In this work, we consider one-particle and two-particle basis states. 
\begin{itemize}
    \item One-particle basis ($|j, \tilde{v}\rangle$): Only the $j$th site is vibronically excited while all other sites are in their vibrational ground states.
    \item Two-particle basis ($\left|j, \tilde{v}; j^\prime, v^\prime\right\rangle$): The $j$th site is vibronically excited with vibronic quantum number $\tilde{v}$ and the $j^\prime$th site is vibrationally excited with vibrational quanta $v^\prime$, all remaining sites (or chromophores) are assumed to be in their vibrational ground states.
\end{itemize}
For example, for a maximum vibrational truncation of $\bar{v}=1$, the system Hamiltonian is constructed from both one-particle and two-particle basis states. The one-particle basis states are $\ket{1,\tilde{0}}$, $\ket{1,\tilde{1}}$, $\ket{2,\tilde{0}}$, and $\ket{2,\tilde{1}}$, whereas the two-particle basis states are $\ket{1,\tilde{0};2,1^\prime}$ and $\ket{2,\tilde{0};1,1^\prime}$. Together, these constitute a six-dimensional multiparticle basis for the system Hamiltonian.

\section{Pure dephasing}

We present our results for the pure dephasing case, i.e. in the absence of any intermolecular coupling $t$. This case serves as a useful benchmark since analytically exact results for the purity decay under pure dephasing dynamics are well established in the literature \cite{Schloss}. Building upon this foundation, we validate our numerical simulations by confirming that they reproduce the known analytical behavior in the pure dephasing limit.

The analytical purity decay is given by:
\begin{align}
\mathcal{P}(t)&=\Tr_{S}\left[\rho_{\rm red}^{2}(t)\right]\nonumber\\
&=\sum_{j=1}^{J}\rho_{jj}^{2}(t)+2\sum_{j<k}|\rho_{jk}(t)|^{2}\label{eq:purity}
\end{align}
For the pure dephasing case, $\rho_{jj}(t)=\rho_{jj}(0)$, since population remains constant due to no intermolecular coupling term. 
$|\rho_{jk}(t)|^{2}=|\rho_{jk}(0)|^{2}|r_{jk}(t)|^{2}$, where $r_{jk}(t)$ is the decoherence factor given by
\begin{align}
    r_{jk}(t)=\prod_{n=1}^{N}r_{jk}^{n}(t)=\exp\left(-\sum_{n=1}^{N}4\frac{|g_{n}|^2}{\omega_{n}^2}(1-\cos\omega_{n}t)\right)
    \label{eq:decohfactor}
\end{align}
\begin{figure}
    \includegraphics[width=\linewidth]{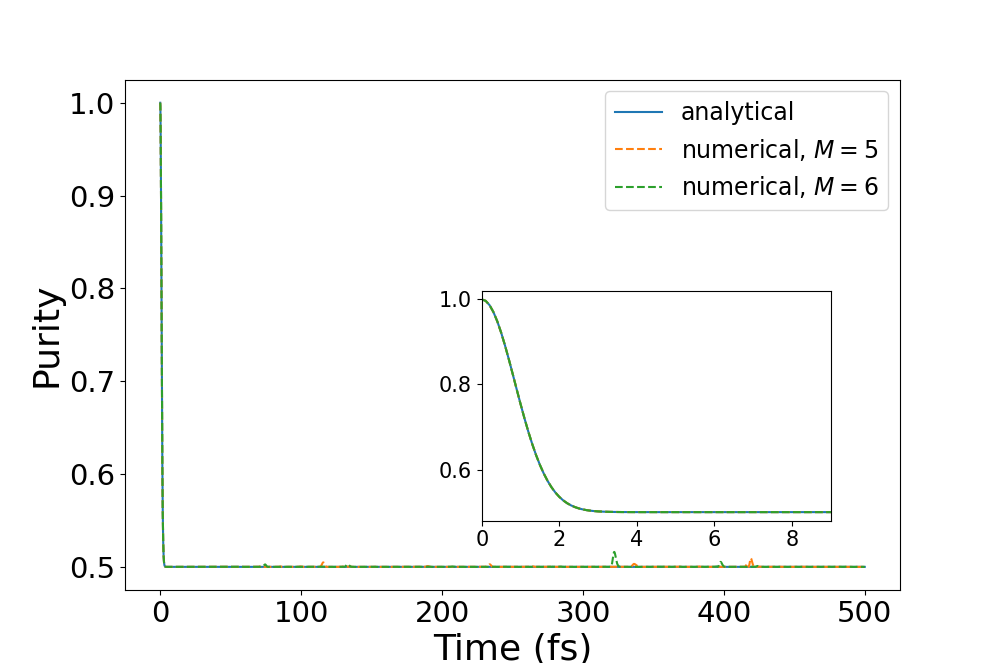}
    \caption{The numerically obtained purity decay for a dimer with no vibrational excitation (maximum vibrational truncation $\bar{v}=0$) is compared with the analytical purity for the pure dephasing case ($t=0$). In the numerical simulations, we consider five bath modes ($N=5$), each represented using five and six Fock states ($M=5,6$). Here we use the high-frequency bath ($\omega_{c}=0.7$ eV).}
    \label{fig:purity_decay_dimer_maxvib0}
\end{figure}
The above expression (eq. (\ref{eq:decohfactor})) is true when we start with coherent superposition of the system basis states and for a zero temperature bath. For a dimer with maximum of zero vibration ($\bar{v}=0$) the initial coherent superposition state is:
\begin{align}
    \ket{\psi_{\rm sys} (0)}=\frac{1}{\sqrt{2}}[1\hspace{0.2cm}1]^{\mathrm{T}}=\sum_{j=1}^{2}\frac{1}{\sqrt{2}}\ket{j}.
\end{align}
In this case of a dimer with no intramolecular vibration, the populations on each site remain constant $\rho_{jj}(t)=\rho_{jj}(0)=1/2$, while the initial coherence is $\rho_{jk}(0)=1/2$. Consequently, using eq. (\ref{eq:purity}) the analytical expression for the purity of the dimer reduces to
\begin{align}
    \mathcal{P}(t)=\frac{1}{2}\left(1+|r_{12}(t)|^2\right)\label{eq:dimer_purity}
\end{align}
The derivation of the decoherence factor has been discussed extensively in the literature, employing approaches such as the Dyson series expansion of the propagator \cite{Schloss} and analytical treatments based on coherent-state evolution \cite{RoBa04}. 

Figure \ref{fig:purity_decay_dimer_maxvib0} presents the pure dephasing dynamics of a dimer without intramolecular vibration. The pure-dephasing calculations were performed using a high-frequency bath with a cut-off frequency of $\omega_{c}=0.7$ eV. To verify numerical convergence, we examined the dependence of the results on the number of bath Fock states. As illustrated in fig. \ref{fig:purity_decay_dimer_maxvib0}, the purity dynamics are well converged with only five and six bath Fock states. 


\section{Derivation of Decoherence Factor}

We further generalize Eq.~(\ref{eq:decohfactor}) to molecular aggregates containing an arbitrary number of chromophores and to system Hilbert spaces of arbitrary dimension arising from the inclusion of multiparticle vibronic basis states. The generalization follows naturally by expressing the decoherence factor solely in terms of the site-dependent sign parameter $s_j$, which completely specifies the system--bath coupling associated with each system basis state. While the analytical purity expression derived above is exact for a dimer (or equivalently a two-level spin-boson system), the corresponding decoherence factor can be extended straightforwardly to arbitrary aggregate sizes because it depends only on the relative system--bath coupling between pairs of basis states.

For the pure dephasing limit ($t=0$), the system Hamiltonian is diagonal in the system basis and each basis state evolves independently under a state-dependent displaced harmonic bath. The coherent-state displacement for the basis state associated with the $j$th site for bath mode $n$ is given by
\begin{align}
\alpha_{j,n}(t)
=
\frac{s_j g_n}{\omega_n}
\left(1-e^{-i\omega_n t}\right),
\end{align}
where $g_n$ and $\omega_n$ denote the coupling strength and frequency of bath mode $n$, respectively, and $s_j=(-1)^{j+1}$ specifies the sign of the system--bath coupling for the corresponding basis state.

The decoherence factor is determined by the overlap between the bath coherent states associated with two different system basis states,
\begin{align}
r_{jk}(t)
=
\langle\Phi_k(t)|\Phi_j(t)\rangle
=
\prod_n
\langle
\alpha_{k,n}(t)
|
\alpha_{j,n}(t)
\rangle.
\end{align}

Using the overlap relation between coherent states,
\begin{align}
\langle\alpha|\beta\rangle
=
\exp\left[
-\frac12|\alpha-\beta|^2
+i\,{\rm Im}(\alpha^*\beta)
\right],
\end{align}
the squared separation between the coherent states generated by basis states associated with $j$th and $k$th sites for mode $n$ becomes
\begin{align}
|\alpha_{j,n}-\alpha_{k,n}|^2
=
\frac{2(s_j-s_k)^2g_n^2}{\omega_n^2}
\left(1-\cos\omega_nt\right).
\end{align}

Substituting this expression into the coherent-state overlap yields the decoherence factor for a single bath mode,
\begin{align}
r_{jk}^{(n)}(t)
=
\exp\left[
-\frac{(s_j-s_k)^2g_n^2}{\omega_n^2}
\left(1-\cos\omega_nt\right)
\right].
\end{align}

Since different bath modes evolve independently, the total decoherence factor is simply the product over all bath modes,
\begin{align}
r_{jk}(t)
=
\prod_n
r_{jk}^{(n)}(t)
=
\exp\left[
-\sum_n
\frac{(s_j-s_k)^2g_n^2}{\omega_n^2}
\left(1-\cos\omega_nt\right)
\right].
\label{eq:decohfac_s}
\end{align}

Equation~(\ref{eq:decohfac_s}) shows that decoherence is governed entirely by the relative system--bath coupling through the quantity $(s_j-s_k)^2$. Consequently, basis states with identical coupling signs ($s_j=s_k$) produce identical bath displacements and therefore remain perfectly coherent, whereas basis states with opposite coupling signs ($s_j=-s_k$) generate different bath coherent states and undergo decoherence. Since Eq.~(\ref{eq:decohfac_s}) depends only on the sign parameter $s_j$, it is independent of the dimension of the system Hilbert space and therefore remains valid for arbitrary molecular aggregate sizes and arbitrary multiparticle Holstein basis sets that include intramolecular vibrational excitations.
\begin{figure}
    \includegraphics[width=\linewidth]{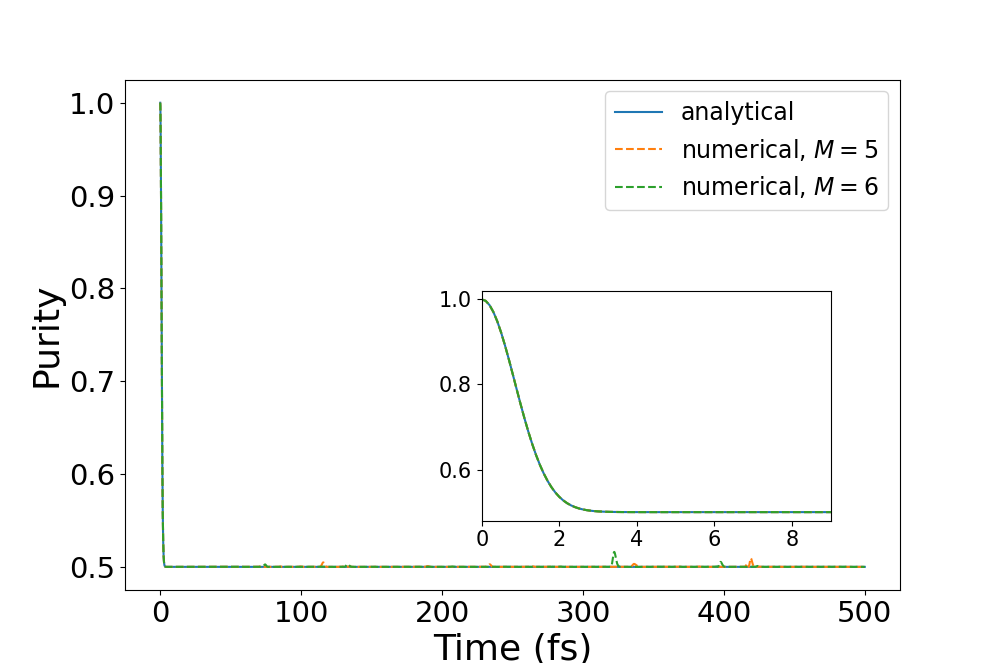}
    \caption{The numerically obtained purity decay for a dimer with a maximum vibrational truncation $\bar{v}=1$) is compared with the analytical purity for the pure dephasing case ($t=0$). In the numerical simulations, we consider five bath modes ($N=5$), each represented using five and six Fock states ($M=5,6$). Here we use the high-frequency bath ($\omega_{c}=0.7$ eV).}
    \label{fig:purity_decay_dimer_maxvib1}
\end{figure}
Thus, the standard decoherence factor obtained for a two-level spin-boson system is recovered as a special case of Eq.~(\ref{eq:decohfac_s}) when the system Hilbert space consists of only two basis states. The present formulation therefore provides a direct extension of the pure-dephasing solution to molecular aggregates of arbitrary size and to multiparticle vibronic Hilbert spaces without requiring any modification of the underlying analytical structure.

Using Eqs.~(\ref{eq:purity}) and (\ref{eq:decohfac_s}), the analytical expression for the purity can now be obtained for a dimer with a maximum vibrational truncation of $\bar{v}=1$. As discussed above, the system Hilbert space is constructed from both one-particle and two-particle basis states. The one-particle basis consists of the states $\ket{1,\tilde{0}}$, $\ket{1,\tilde{1}}$, $\ket{2,\tilde{0}}$, and $\ket{2,\tilde{1}}$, while the two-particle basis consists of $\ket{1,\tilde{0};2,1^\prime}$ and $\ket{2,\tilde{0};1,1^\prime}$. Together, these span a six-dimensional multiparticle Hilbert space. Accordingly, the initial coherent superposition state is chosen as an equal superposition of all six basis states,
\begin{align}
    \ket{\psi_{\rm sys}(0)}
    =
    \frac{1}{\sqrt{6}}
    \sum_{j=1}^{6}
    \ket{j}.
\end{align}
Therefore, using Eqs. (\ref{eq:purity}) and (\ref{eq:decohfac_s}), one could derive the analytical purity for the case of dimer with maximum of one vibration, which also reduces to Eq. (\ref{eq:dimer_purity}).  

Figure \ref{fig:purity_decay_dimer_maxvib1} shows the numerical purity decay along with the analytical one for this case of maximum vibrational truncation of $\bar{v}=1$ and they are exactly on top of each other. 

\section{Convergence Tests}

\begin{figure}[H]
    \includegraphics[width=\linewidth]{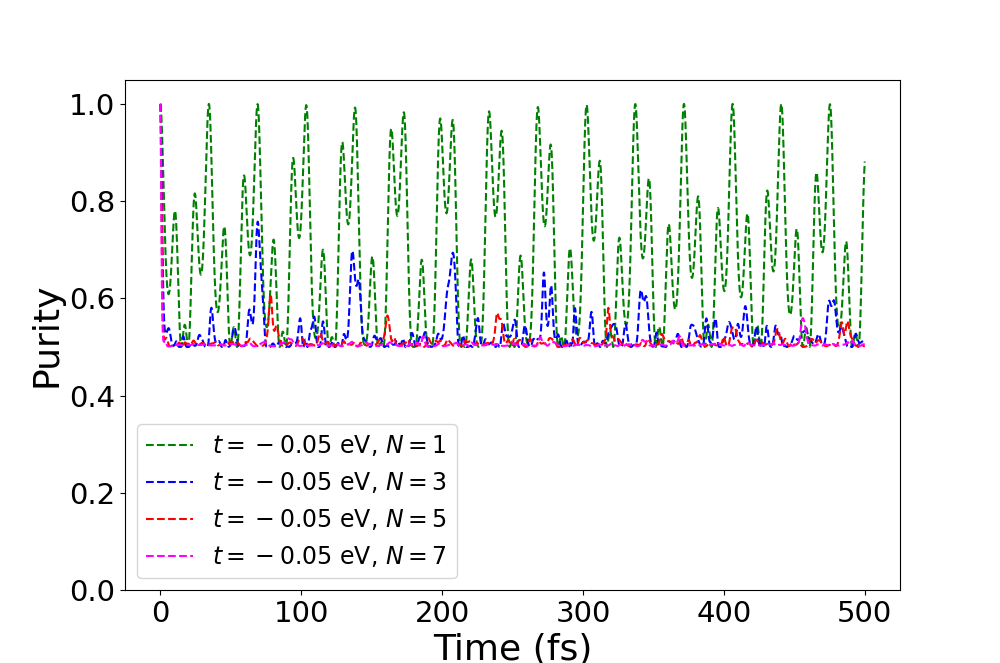}
    \caption{Convergence test for a purely electronic J-dimer ($t=-0.05$) eV coupled to a high-frequency bath ($\omega_{c}=0.7$ eV).}
    \label{fig:convergence_J}
\end{figure}

\begin{figure}[H]
    \includegraphics[width=\linewidth]{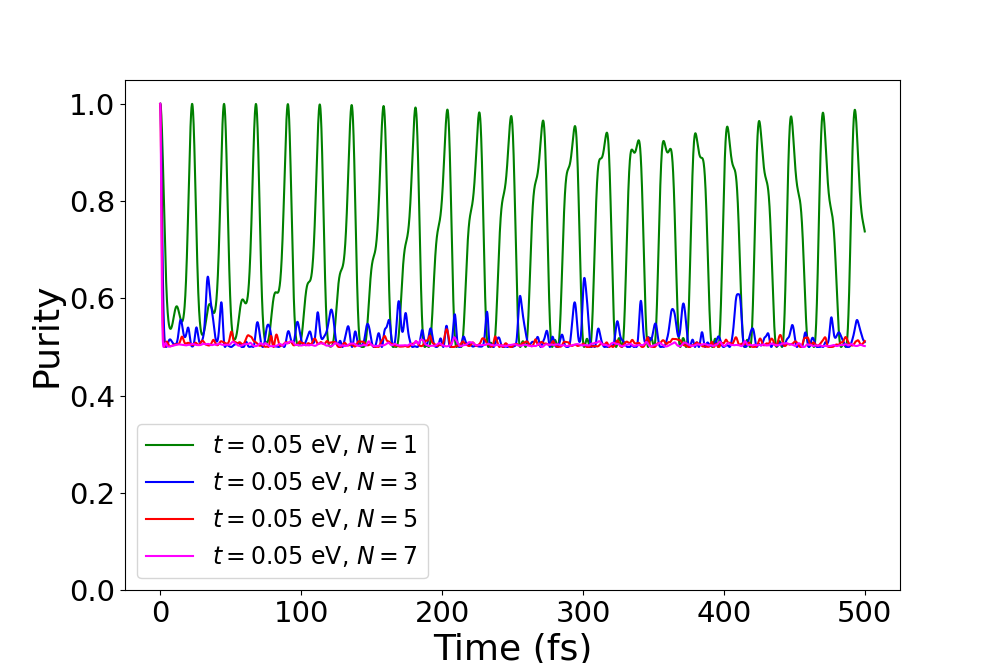}
    \caption{Convergence test for a purely electronic H-dimer ($t=0.05$) eV coupled to a high-frequency bath ($\omega_{c}=0.7$ eV).}
    \label{fig:convergence_H}
\end{figure}

To verify the numerical convergence of the purity dynamics with respect to the number of bath modes, we consider H- and J-dimers ($t=\pm 0.05$ eV) in the purely electronic limit ($\bar{v}=0$) coupled to a high-frequency bath ($\omega_{c}=0.7$ eV). The calculations are performed for $N=1,3,5$ and $7$ bath modes, with each mode truncated to $M=3$ Fock states. The results demonstrate excellent convergence with increasing bath discretization. Figures \ref{fig:convergence_J} and \ref{fig:convergence_H} shows the convergence of the purity for J- and H-dimers with increasing bath discretization.
Figures \ref{fig:convergence_J} and \ref{fig:convergence_H} demonstrates that the purity dynamics are well converged with five bath oscillators. 

\section{Larger Molecular Aggregates: Pentamer}
\label{sec:pentamer}

\begin{figure}[H]
    \includegraphics[width=\linewidth]{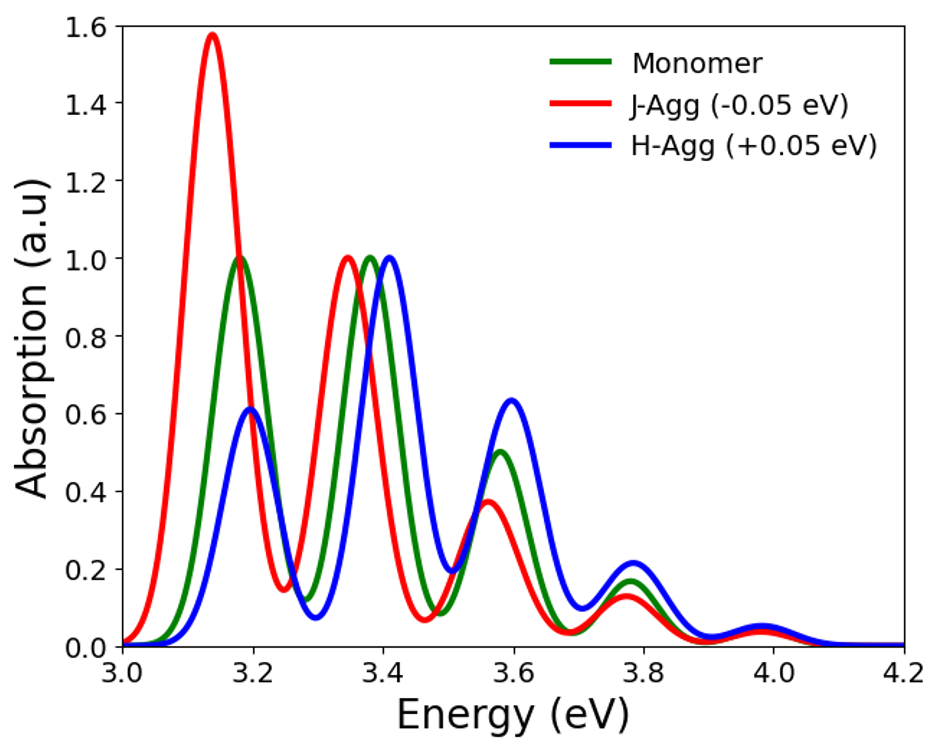}
    \caption{Vibronic absorption spectra for a pentamer showing the redistribution of oscillator strength among the $0-0$, $0-1$, and higher order transitions for H- and J-aggregates.}
    \label{fig:pentamer1}
\end{figure}

Figure \ref{fig:pentamer1} presents the calculated absorption spectra of the pentamer, including intramolecular vibronic coupling, for both H- and J-aggregate configurations, together with the corresponding monomer spectrum. The characteristic spectral signatures of H- and J-aggregates are preserved upon increasing the aggregate size. For the J-aggregate ($t=-0.05$ eV), the lowest-energy $0–0$ transition exhibits a pronounced enhancement in absorption intensity relative to the monomer, accompanied by a redshift of the spectral maximum. Conversely, the H-aggregate ($t=+0.05$ eV) displays a blueshifted $0–0$ transition with suppressed intensity. The redistribution of oscillator strength across the vibronic progression reflects the different energetic arrangements of the optically bright states in H- and J-aggregates.
To examine the influence of the environmental spectral window on decoherence, we next consider the pentamer in the purely electronic limit, neglecting intramolecular vibronic coupling. The dynamics are initialized in the corresponding optically bright excitonic eigenstates, and the resulting subsystem purity is evaluated for two Ohmic bath cutoff frequencies, $\omega_{c}=0.052$ eV and $\omega_{c}=0.7$ eV.
\begin{figure}[H]
    \includegraphics[width=\linewidth]{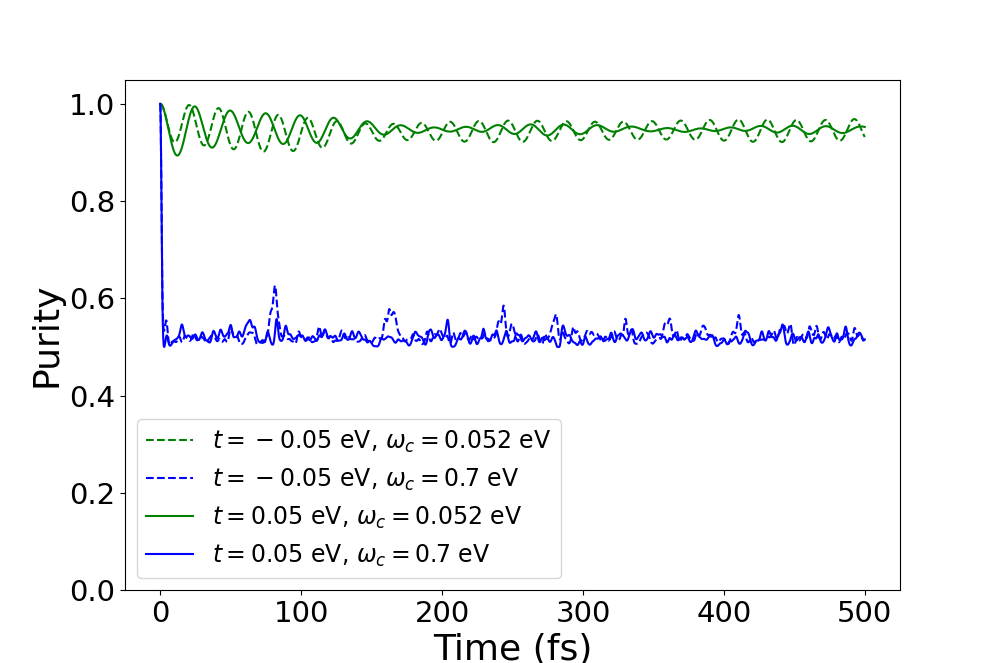}
    \caption{Purity dynamics of the purely electronic pentamer for the weak-coupling regime ($t=\pm 0.05$ eV) corresponding to $\omega_{c}=0.052$ eV and $\omega_{c}=0.7$ eV.}
    \label{fig:pentamer2}
\end{figure}

\bibliography{references}